\documentclass[final,5p,times,preprint]{elsarticle}
\usepackage[utf8]{inputenc}
\usepackage{amsfonts}
\usepackage{amsmath}
\usepackage{amssymb}
\usepackage{bm}
\usepackage{lineno,hyperref}
\usepackage{comment}
\usepackage{xcolor}
\usepackage{booktabs}
\usepackage{multirow}%divide cells in table
\modulolinenumbers[5]
\usepackage[normalem]{ulem}

\journal{Physics Letters B}

\biboptions{comma,sort&compress}

\begin{document}

\begin{frontmatter}

%\title{Transfer to continuum calculations for the $^{11}$Li$(p,pn){^{10}}$Li quasifree reaction}
%\title{A model for $(p,pn)$ reactions with Borromean nuclei: the $^{11}$Li$(p,pn){^{10}}$Li case}
\title{Binding-energy independence of reduced spectroscopic strengths derived from $(p,2p)$ and $(p,pn)$ reactions with nitrogen and oxygen isotopes}


\author[FAMN]{M. G\'omez-Ramos\corref{mail}}
\cortext[mail]{Corresponding author}
\ead{mgomez40@us.es}

\author[FAMN]{A. M. Moro}

\address[FAMN]{Departamento de F\'{\i}sica At\'omica, Molecular y Nuclear, Facultad de F\'{\i}sica, Universidad de Sevilla, Apartado 1065, E-41080 Sevilla, Spain}

\begin{abstract}
%\begin{description}
A campaign of intermediate energy (300-450~MeV/u) proton-induced nucleon knockout measurements in inverse kinematics has been recently undertaken at  the R$^3$B/LAND setup at GSI. We present a systematic theoretical analysis of these data with the aim of studying the quenching of the single-particle strengths and its binding-energy dependence. For that, the measured semi-inclusive $(p,2p)$ and $(p,pn)$ cross sections  are compared with theoretical predictions based on single-particle cross sections derived from a novel coupled-channels formalism and 
shell-model spectroscopic factors. A systematic reduction of about 20-30\% is found, with a very limited dependence on proton-neutron asymmetry. 
%\end{description}
\end{abstract}

\end{frontmatter}

%\tableofcontents
%-------------------------------------------------
\section*{Introduction}
%-------------------------------------------------
The atomic nucleus is a complicated  many-body system of strongly correlated fermions. The idea, first proposed by Mayer \cite{May69} and Haxel {\it et al.} \cite{Hax49},  of treating the motion of the nucleons as independent particles moving in a mean-field potential led to a  remarkably simple picture, the independent particle shell-model (IPM), whose most notable  success is the explanation  of magic numbers in terms of main-shell closure.

Not surprisingly, this appealing but highly simplified description of the nucleus has limitations. Beyond mean-field effects lead to deviations of the IPM which  manifest as a 
%can be conveniently casted in the form of 
fragmentation of the single-particle levels and the subsequent depletion of their occupancies. This effect is usually quantified making use of the spectroscopic factor (SF), which is the norm of the overlap between the $A$ and $A-1$ many-body wave functions \cite{Jen11}. The SF is a measure of how well a nucleus $A$ can be described by a single-particle nucleon attached to a $A-1$ {\it core}. Since the mean-field potential defining the single-particle basis is not unique, the SF are not unique either \citep{Fur10}. Still, they are useful quantities to describe the behavior of nucleons in the nucleus.

Within these model-dependence constraints, SF 
or, more generally, overlap functions are essential inputs of reaction calculations. Therefore, information about the SF can in principle be obtained by comparing experimental cross sections with theoretical predictions. 
In general, it has been found that these theoretical cross sections tend to overestimate experimental ones, and it is common to define a reduction or ``quenching'' factor $R_s=\sigma_\mathrm{exp}/\sigma_\mathrm{th}$. Systematic $(e,e'p)$ studies on stable nuclei, as those performed at NIKHEF \cite{Lap93}, suggest that the spectroscopic factor of protons in valence orbits are reduced by 30-40\% with respect to the IPM prediction. Similar reductions have been found in systematics for different transfer reactions \citep{Kay13}.

These studies have been later extended to more asymmetric systems, using heavy-ion knockout reaction experiments at medium energies up to 100 MeV in which a fast-moving projectile nucleus collides with a stable composite nucleus  (such as $^{9}$Be or $^{12}$C) losing a nucleon. The analysis of these reactions with the eikonal reaction theory \cite{Gad08}, assuming spectroscopic factors from shell-model calculations shows also a sizable quenching  but, most notably, with a strong isospin dependence, which manifests as a dependence on the difference between separation energies $\Delta S=S_{p(n)}-S_{n(p)}$, for proton (neutron) removal. In particular, it is found that $R_s$  is close to unity for the removal of weakly bound nucleons, whereas it is much smaller than 1 for deeply bound ones. This  has been interpreted as an indication of additional  correlations, which cannot be described properly by the shell model \citep{Bar09}.

However, this marked  dependence on $\Delta S$ does not seem to be supported by  the results obtained with transfer reactions  \citep{Lee10,Fla13,Fla18}. Furthermore, state-of-the-art  {\it ab-initio} calculations \cite{Jen11} display in fact some  dependence, in qualitative agreement with knockout results, but to a much more modest degree.

It is worth noting that the theoretical cross section ($\sigma_\mathrm{th}$)  that enters the definition of $R_s$  involves both the spectroscopic factors and the description of the reaction mechanism, through the single-particle cross section $\sigma_{sp}$ ($\sigma_\mathrm{th}= \mathrm{SF} \times \sigma_{sp}$). The spectroscopic factors are usually obtained from shell-model calculations, for which multiple predictions exist, introducing a measure of uncertainty in the $R_s$. The $R_s$ is also dependent on the description of the reaction mechanism and, as such, the different behavior of the $R_s$ values extracted from transfer and knockout experiments might be actually due to inadequacies in the reaction models employed in either of these analyses. In particular, the validity of the sudden approximation, which is commonly assumed in the analysis of knockout experiments, has been put into question for the removal of deeply-bound nucleons  \cite{Fla13,Sun16}. Conversely, the analysis of transfer reactions is known to be affected by significant uncertainties \cite{Lov15,Nun11}.

To shed light on this complicated scenario, several experimental facilities have undertaken systematic studies of $(p,pN)$ reactions at intermediate energies (several hundreds of MeV per nucleon), using radioactive beams on hydrogen targets \cite{Dia18,Ata18,Pan16,Kaw18}. These results share some similitudes with heavy-ion induced knockout reactions but with two main differences. First, they are expected to probe deeper portions of the nuclear densities \citep{Ryc11,Aum13} and, second, the final state can be fully determined provided that the three outgoing fragments (the residual nucleus and the two outgoing nucleons) are measured. The study from \cite{Kaw18} spans a series of $(p,2p)$ fully exclusive measurements on several stable and unstable oxygen isotopes from the Radioactive Isotope Beam Factory (RIBF) at RIKEN  and the Research Center for Nuclear Physics
(RCNP) at Osaka. Reduction factors were obtained by comparing the measured cross sections to DWIA calculations using shell-model spectroscopic factors. The derived $R_s$ values fluctuate from 0.5 to 0.7 and show no evidence of significant $\Delta S$ dependence. The measurements of  \cite{Dia18,Ata18} were performed at the R$^3$B-LAND setup at GSI, and correspond to semi-inclusive  (summed over the bound states of the residual nucleus) cross sections.
The work of \cite{Ata18} analyzes $(p,2p)$ data for five selected oxygen isotopes ($^{A}$O($p$,$2p$)$^{A-1}$N, with $A=$14, 16, 17, 21 and 23), which were assumed to be well closed shell nuclei. The experimental cross sections for these nuclei were compared with eikonal DWIA calculations \cite{Aum13}, leading to an average value for the reduction factor of $R_s \sim 0.66$, with a rather weak $\Delta$S dependence. The work of \cite{Dia18} comprises several oxygen and nitrogen isotopes, and both $(p,2p)$ and $(p,pn)$ cases. The data were analyzed with the Alt-Grassenber-Sandas (AGS) formulation of the Faddeev three-body formalism \cite{AGS}. Comparison of the experimental cross sections with these calculations leads to $R_s$ values significantly smaller than unity, ranging from $R_s=0.67$ for $^{21}$N$(p,pn)$ to $R_s=0.32$ for $^{21}$N$(p,2p)$. Since the theoretical analysis of these two works were performed with different reaction formalisms and different structure and potential inputs, a comparison between these results needs to be done with caution.

Given the differences between the results of Refs.~\cite{Dia18} and \cite{Ata18}, in this work we present a joint analysis of the full set of experimental data from these works, employing a common reaction formalism, namely, the transfer-to-the-continuum (TC) method \cite{Mor15} and shell-model spectroscopic factors derived from the same NN effective interaction.  
With this analysis, we show that a rather consistent picture  can be obtained from the full set of experimental measurements reported in \citep{Dia18,Ata18}.

\medskip %-------------------------------------------------
\section*{Theoretical approach}
%-------------------------------------------------
The process under study is of the form $p+A \rightarrow p +N+B$ with $N=p$ or $N=n$ for $(p,2p)$ or $(p,pn)$ reactions, respectively. This process is described with the the transfer to the continuum (TC) reaction formalism \cite{Mor15}, which is based on the prior form transition amplitude for the process $A(p,pN)B$:
\begin{equation}
T_{if}^{nljm}=
\left\langle
\Psi_f^{3b(-)}
|V_{pN}+V_{pB}-U_{pA}|
\chi_{0,\bm{K}_0}^{(+)}\varphi^{nlj m}
\right\rangle,
\end{equation}
where $\varphi^{nlj m}$ is the bound nucleon wave function, $\chi_{0,\bm{K}_0}^{(+)}$ is the distorted wave between the incoming proton with momentum $\bm{K}_0$ and the target and $\Psi_f^{3b(-)}$ is the final 3-body wave function ($p,N$ and residual core $B$), with $V_{xy}$, $U_{xy}$ being the binary interactions between $x$ and $y$.
The final three-body wave function describing the $p+N+A$ system is expanded in terms of $p-N$ states, for a wide range of relative energies and as many partial waves needed to achieve convergence of the calculated observables. A procedure of continuum discretization, similar to that used in the continuum-discretized coupled-channels (CDCC) method \cite{Aus87}, is used to make the sum discrete and finite. 
\begin{equation}
\Psi_f^{3b(-)}(\vec r,\vec R) \approx\sum_{i j'\pi} \phi_i^{j'\pi}(k_i,\vec r)\chi_i^{j'\pi}(\vec K_i,\vec R),
\end{equation}
where $i$ is the index the discretized state $ \phi_i^{j'\pi}$ (with an associated center of mass momentum between the $(pN)$ ensemble and $B$: $\vec K_{i}$ and $p-N$ relative momentum $k_i$), with angular momentum and parity between $p$ and $N$ $j'^\pi$.
The resultant expression of the transition amplitude is then  formally similar to that used in the coupled-channels Born approximation (CCBA) method for transfer reactions: 
\begin{align}
T_{if} &\approx \sum_{i j'\pi}\left\langle
\phi_i^{j'\pi} \chi_i^{j'\pi}
|V_{pN}  %\nonumber \\ 
 + V_{pB}-U_{pA}|
\chi_{0,\bm{K}_0}^{(+)}\varphi^{nlj m}
\right\rangle.
\end{align}
As such, it can be computed with standard coupled-channels codes. In here, we use a modified version of the code {\sc fresco} \cite{fresco}, which incorporates the Reid93 NN interaction, and the relativistic kinematics corrections discussed in \cite{Mor15}. 

%Potentials
Important ingredients of the calculations are the distorting potentials describing the relative motion of the incident and outgoing nucleons with respect to projectile and residual nuclei, respectively. In particular, the imaginary part of these potentials accounts for the absorption and re-scattering effects of the incident and outgoing nucleons. Two sets of distorting potentials are considered in this work. One of them is the phenomenological Dirac parametrization based on the EDAD2 parameter set \citep{Ham90,Coo93}. The other are microscopic optical potentials generated  by folding the Paris-Hamburg (PH)  $g$-matrix  NN effective interaction \cite{Ger83,Rik84} with the ground-state density of the corresponding composite nucleus, obtained from a Hartree-Fock (HF) calculation using the Skyrme SkX interaction. Both potentials are energy dependent.  For the incident channel, the potential is evaluated at the incident energy ($E_\mathrm{lab}$). For the exit channel, the choice is less clear, because the outgoing nucleons will emerge with a broad range of energies. For a pure quasi-free collision between the incident proton and the knocked out nucleon, one expects an average value of about $E_\mathrm{lab}/2$ for each nucleon, and so the outgoing optical potentials were evaluated at this average energy. 

The overlap between nuclei $A$ and $A-1$ has been approximated by a single-particle wave function, generated as the eigenstate of a Woods-Saxon potential,  and normalized to the shell-model spectroscopic factor. Following \citep{Gad08}, a diffuseness of $a=0.7$~fm was adopted in all cases and the radius and depth of the potential were adjusted in order to reproduce 
the mean square radius of the aforementioned Hartree-Fock calculation and the experimental separation energy. 
A spin-orbit term with the same radius and diffuseness and depth $V_{so}=6$ MeV is also included. 
The dependence of the obtained cross sections on the choice of the nucleon-nucleon interaction used in the Hartree-Fock calculations  has been tested in some cases by using the Skyrme Skm$^*$ interaction \citep{Bar82} instead of SkX. The results are discussed below.

%The spectroscopic factors assigned to each of the involved states of the residual \textit{core} nucleus have been computed using the WBT interaction of  Warburton and Brown \cite{WBT}, assuming a $psd$ configuration space with $n$ particle-$n$ hole excitations, taking $n$ as the minimum value required by Pauli principle. An exception is made for $^{16}$O, where $n=0+2$ was considered.

The spectroscopic factors assigned to each of the involved states of the residual \textit{core} nucleus have been computed using the WBT interaction of  Warburton and Brown \cite{WBT}, assuming a $psd$ configuration space, with $n$ particle-$n$ hole excitations, taking $n$ as the minimum value required to produce a non-zero overlap between the projectile nucleus in its ground state and the residual \textit{core}. An exception is made for $^{16}$O, where $n=0+2$ was considered. To test the dependence on the spectroscopic factors, another prescription for them has also been used, using Cohen-Kurath interaction \citep{Coh65} for nuclei $^{13-16}$O, which do not involve the $sd$ shell, and PSDMK interaction \citep{Mil75} for the rest of nuclei.
The factor $\left(A/(A-1)\right)^N$ considered in previous works \citep{Die74,Tos14} has been included. Note that this factor has been strictly derived for a harmonic oscillator model, so it might not be suitable for all the nuclei considered.

\medskip
%--------------------------------------------------------------
\section*{Comparison to $(p,2p)$ and $(p,pn)$ data} 
\begin{figure}[tb]
\begin{center}
% {\def\svgwidth{0.9\columnwidth}{\input{zrcoor.pdf_tex}} \par}
 {\centering \resizebox*{0.85\columnwidth}{!}{\includegraphics{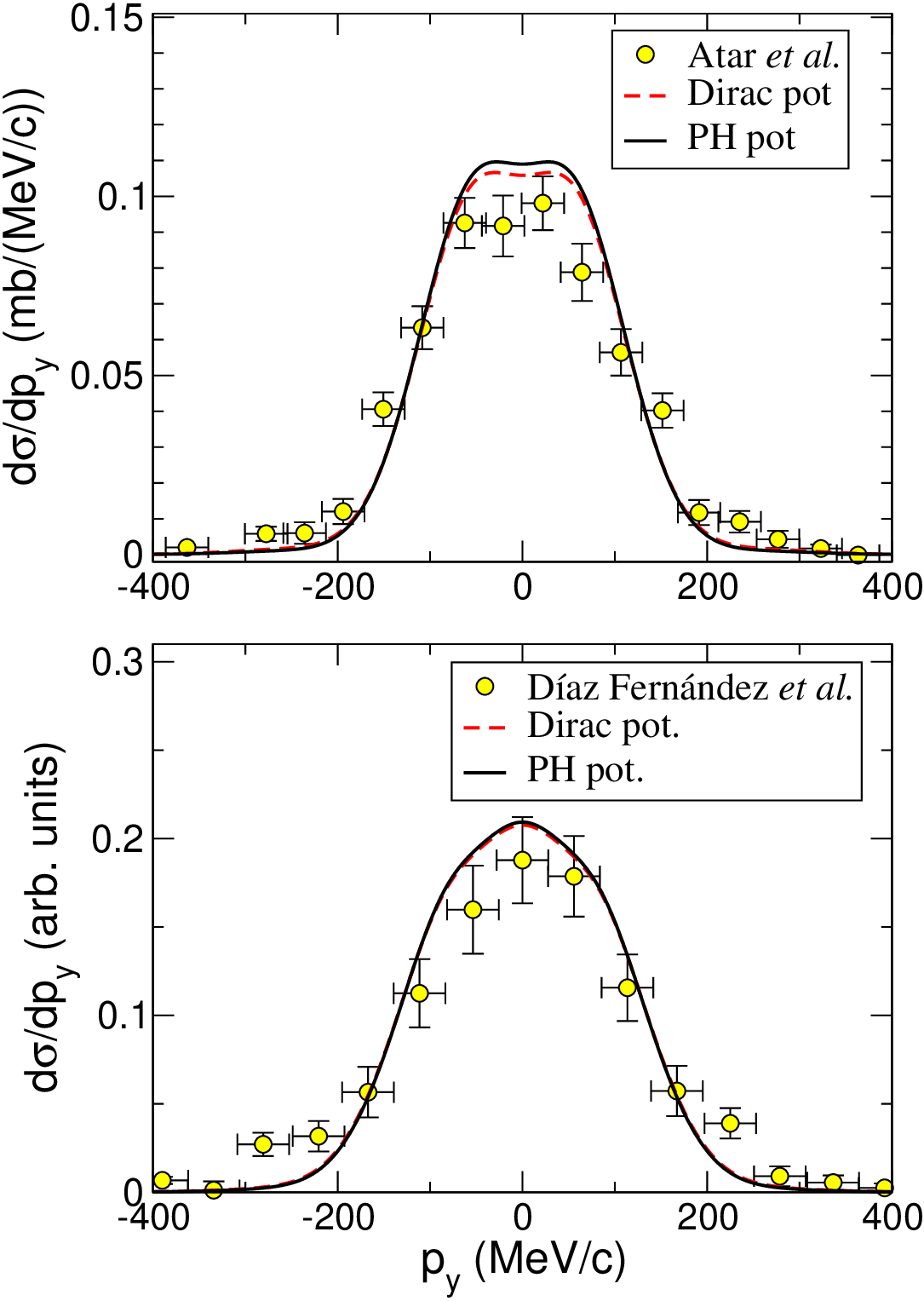}} \par}
\caption{\label{fig:dsdpx}(Color online)Transversal momentum distribution for $^{16}$O$(p,2p)$ (top) and $^{22}$O$(p,pn)$ (bottom). Experimental data are taken from Refs.~\citep{Ata18} and \cite{Dia18}, respectively. The solid and dashed  lines correspond to the present calculations using Paris-Hamburg and Dirac potentials (see text), respectively. For the top figure, calculations have been rescaled by the quenching factors required to reproduce the integrated $(p,2p)$ cross section (0.74 for Dirac potentials and 0.78 for PH potentials), while for the bottom figure, since experimental data are presented in arbitrary units, calculations have been rescaled to give the same total integral.}
\end{center}
\end{figure}

%---------------------------------------------------------------
Although the quenching factors quoted in this work are obtained from the ratio of integrated $(p,pN)$ cross sections,  a more detailed comparison between theory and experiment can be done for the measured momentum distributions. This comparison provides a more stringent test of the validity of the reaction theory. As representative examples, in Fig.~\ref{fig:dsdpx} we compare the experimental and calculated transverse momentum distributions  for $^{16}$O$(p,2p)$ from \citep{Ata18} and for $^{22}$O$(p,pn)$ from \citep{Dia18}, respectively. As in the other cases discussed below, the data are inclusive with respect to the populated states of the residual nucleus. Therefore,  the calculations correspond to the sum over the bound states of this nucleus.  The solid and dashed lines correspond, respectively, to the PH and Dirac potentials. 

 In the $^{16}$O$(p,2p)$ case, the  theoretical results have been scaled by the extracted quenching factors ($R_s=0.74$ and 0.78, for the Dirac and PH potentials, respectively). For the  $^{22}$O$(p,pn)$ case, since experimental data were given in arbitrary units, theoretical distributions have been rescaled to reproduce the total integral of the given data. It can be seen that the choice of the distorting potentials has a negligible effect on the shape of the momentum distributions, although the corresponding $R_s$ values differ by 5-6\% (see Table \ref{tab:rs3}). Theoretical momentum distributions agree reasonably well with experimental data, although they tend to overestimate the peak at $p_x=0$ and seem to be  narrower in the tail region. In a recent comparison between the Transfer to the Continuum and the DWIA formalisms \citep{Yos18} it was found that the neglect of the energy dependence of the potentials produced distributions which were slightly narrower than those which included energy-dependent potentials. Since  in the TC method this energy dependence is not taken into account, this can be pointed as a possible cause for the narrowness of the distributions. 

\begin{comment}
\begin{figure}[tb]
\begin{center}
\begin{minipage}[t]{\columnwidth}
 {\centering \resizebox*{0.85\columnwidth}
 {!}{\includegraphics{o16p2p_dsdpy.eps}} \par}
 \end{minipage}
\begin{minipage}[t]{\columnwidth}
 {\centering \resizebox*{0.85\columnwidth}{!}{\includegraphics{dsdpx_paper_220.eps}} \par}
\end{minipage}
\caption{\label{fig:dsdpx}(Color online)Transversal momentum distribution for $^{16}$O$(p,2p)$ (top) and $^{22}$O$(p,pn)$ (bottom). Experimental data are taken from Refs.~\citep{Ata18} and \cite{Dia18}, respectively. The solid and dashed  lines correspond to Transfer to the Continuum calculations using Paris-Hamburg and Dirac potentials (see text), respectively. For the top figure, calculations have been rescaled by the quenching factors shown in Table \ref{tab:rs3}, while for the bottom figure, since experimental data are presented in arbitrary units, calculations have been rescaled to give the same total integral. }
\end{center}
\end{figure}
\end{comment}

\begin{figure}[tb]
\begin{center}
% {\def\svgwidth{0.9\columnwidth}{\input{zrcoor.pdf_tex}} \par}
 {\centering \resizebox*{\columnwidth}{!}{\includegraphics{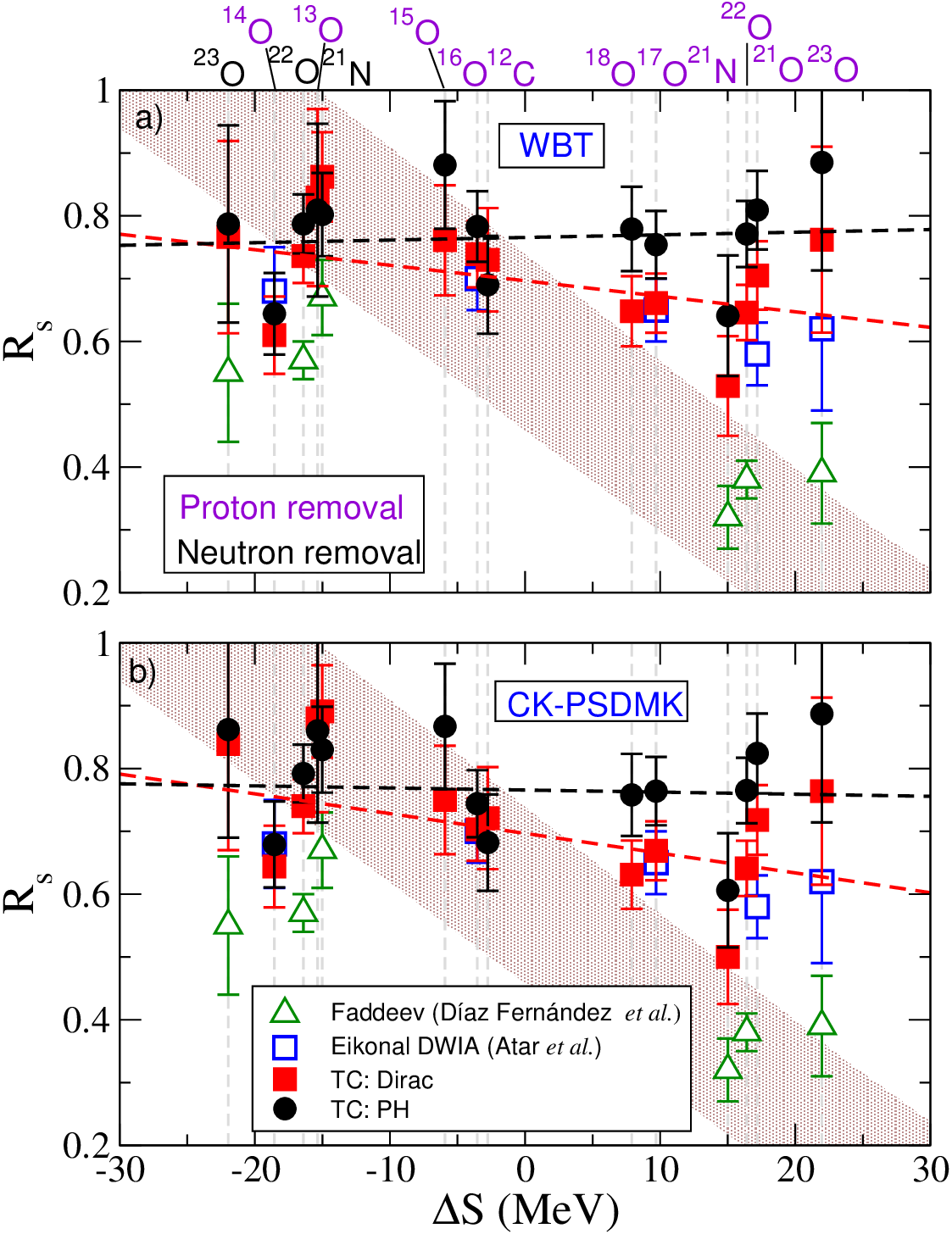}} \par}
\caption{\label{fig:Rs} Reduction factors obtained for different $(p,pn)$ and $(p,2p)$ reactions as a function of $\Delta S$ (see text). Red squares and black circles correspond to calculations using Dirac and Paris-Hamburg potentials, respectively. A linear fit of each set is presented in the red and black dashed lines respectively. Blue empty squares correspond to the analysis performed in \citep{Ata18} and green empty triangles to the one in \citep{Dia18}. The brown band indicates the trend found for nucleon knockout reactions with composite nuclei  \citep{Gad08}. The top panel shows calculations in which the SF have been computed using WBT interaction, while the bottom one shows calculations using Cohen-Kurath interaction for reactions on nuclei $^{13-16}$O and PSDMK interaction for the rest.}
\end{center}
\end{figure}

Unfortunately, this procedure could not be applied to all the reactions  considered in this work because, for some of them, momentum distributions have not been published and, for others, they have a very limited statistics. Consequently, to extract the $R_s$ values we have considered the ratio between the integrated cross sections. The results of these calculations, based on the WBT shell-model interaction, are listed in Table \ref{tab:rs3}
. In it, the second to fifth columns correspond to the sum of the spectroscopic factors corresponding to all bound states of the residual \textit{core} which couple to the indicated single-particle states to produce the bound state of the projectile nucleus. The sixth column indicates the average single particle cross section $\sigma_\mathrm{sp}$, computed using Dirac (upper value) and PH (lower value) potentials. The seventh column shows the theoretical cross section $\sigma_\mathrm{th}=\sum C^2S \sigma_\mathrm{sp}$, and the eighth corresponds to the experimental cross sections $\sigma_\mathrm{exp}$ from \citep{Ata18,Dia18,Pan16}. For $^{22}$O,$^{23}$O$(p,2p)$, analyses giving different but compatible cross sections have been given in \citep{Ata18} and \citep{Dia18}. In these cases we have considered the values from \citep{Dia18}. Finally, the quenching factor $R_s=\sigma_\mathrm{exp}/\sigma_\mathrm{th}$ is shown in the last column. The extracted reduction factors are also shown in Fig.~\ref{fig:Rs} as a function of the difference of the separation energies $\Delta S$. Red squares and black circles correspond, respectively, to the Dirac and PH potentials. The $R_s$ factors obtained in Refs.~\citep{Ata18} 
%using eikonal DWIA reaction theory \citep{Aum13} 
and \citep{Dia18}
%using the Faddeev/AGS formalism 
are also shown in blue empty squares and green empty triangles, respectively, for the sake of comparison. Results are presented with error bars propagated from the errors for the experimental cross sections. 

\begin{table*}[b]%The best place to locate the table environment is directly after its first reference in text

\begin{tabular}{l|ccccccc|c}
Reaction  &
$\sum C^2S_{1p_{3/2}} $ &
$\sum C^2S_{1p_{1/2}} $ &
$\sum C^2S_{1d_{5/2}} $ &
$\sum C^2S_{2s_{1/2}} $ &
$\sigma_\mathrm{sp}$  &
$\sigma_\mathrm{th}$  &
$\sigma_\mathrm{exp}$  &
$R_s$  \\
\hline

\multirow{2}{*}{$^{13}$O$(p,2p)$}&\multirow{2}{*}{--}&\multirow{2}{*}{0.66} &\multirow{2}{*}{--} & \multirow{2}{*}{--} & 10.562& 6.975 &\multirow{2}{*}{5.78(0.91)[0.37]}&0.83(14)\\
&&&& & 10.813&7.140&&0.81(14)\\[1mm]
%\colrule
\multirow{2}{*}{$^{14}$O$(p,2p)$}&\multirow{2}{*}{--}&\multirow{2}{*}{1.97} &\multirow{2}{*}{--} & \multirow{2}{*}{--} & 8.509& 16.769 &\multirow{2}{*}{10.23(0.80)[0.65]}&0.61(6)\\
&&&& & 8.065&15.895&&0.64(6)\\[1mm]
%\colrule

\multirow{2}{*}{$^{15}$O$(p,2p)$}&\multirow{2}{*}{1.94}&\multirow{2}{*}{1.60} &\multirow{2}{*}{--} & \multirow{2}{*}{--} & 7.026& 24.856 &\multirow{2}{*}{18.92(1.82)[1.20]}&0.76(9)\\
&&&& & 6.072&21.481&&0.88(10)\\[1mm]
%\colrule
\multirow{2}{*}{$^{16}$O$(p,2p)$}&\multirow{2}{*}{4.09}&\multirow{2}{*}{2.00} &\multirow{2}{*}{--} & \multirow{2}{*}{--} & 5.965& 36.308 &\multirow{2}{*}{26.84(0.90)[1.70]}&0.74(5)\\
&&&& & 5.631&34.279&&0.78(6)\\[1mm]
%\colrule
\multirow{2}{*}{$^{17}$O$(p,2p)$}&\multirow{2}{*}{--}&\multirow{2}{*}{2.07} &\multirow{2}{*}{--} & \multirow{2}{*}{} & 5.777& 11.944 &\multirow{2}{*}{7.90(0.26)[0.50]}&0.66(5)\\
&&&& & 5.064&10.471&&0.75(5)\\[1mm]
%\colrule
\multirow{2}{*}{$^{18}$O$(p,2p)$}&\multirow{2}{*}{3.40}&\multirow{2}{*}{2.04} &\multirow{2}{*}{--} & \multirow{2}{*}{--} & 5.051& 27.488 &\multirow{2}{*}{17.80(1.04)[1.13]}&0.65(6)\\
&&&& & 4.201&22.863&&0.78(7)\\[1mm]
%\colrule
\multirow{2}{*}{$^{21}$O$(p,2p)$}&\multirow{2}{*}{--}&\multirow{2}{*}{1.88} &\multirow{2}{*}{--} & \multirow{2}{*}{--} & 4.008& 7.532 &\multirow{2}{*}{5.31(0.23)[0.34]}&0.71(5)\\
&&&& & 3.493&6.5656&&0.81(6)\\[1mm]
%\colrule
\multirow{2}{*}{$^{21}$N$(p,2p)$}&\multirow{2}{*}{0.33}&\multirow{2}{*}{0.72} &\multirow{2}{*}{--} & \multirow{2}{*}{--} & 4.118& 4.290 &\multirow{2}{*}{2.27(0.34)}&0.53(8)\\
&&&& & 3.398&3.540&&0.64(10)\\[1mm]
%\colrule
\multirow{2}{*}{$^{21}$N$(p,pn)$}&\multirow{2}{*}{--}&\multirow{2}{*}{--} &\multirow{2}{*}{4.95} & \multirow{2}{*}{0.65} &10.059 & 56.274 &\multirow{2}{*}{48.52(4.04)}&0.86(7)\\
&&&& & 10.809&60.471&&0.80(7)\\[1mm]
%\colrule
\multirow{2}{*}{$^{22}$O$(p,2p)$}&\multirow{2}{*}{0.73}&\multirow{2}{*}{1.87} &\multirow{2}{*}{--} & \multirow{2}{*}{--} & 3.533& 9.175 &\multirow{2}{*}{6.01(0.41)}&0.65(4)\\
&&&& & 2.962&7.693&&0.77(5)\\[1mm]

\multirow{2}{*}{$^{22}$O$(p,pn)$}&\multirow{2}{*}{--}&\multirow{2}{*}{--} &\multirow{2}{*}{5.89} & \multirow{2}{*}{0.25} & 8.690 & 53.349 &\multirow{2}{*}{39.24(2.34)}&0.74(4)\\
&&&& & 8.122&49.865&&0.79(5)\\[1mm]

\multirow{2}{*}{$^{23}$O$(p,2p)$}&\multirow{2}{*}{--}&\multirow{2}{*}{1.99} &\multirow{2}{*}{--} & \multirow{2}{*}{--} & 3.302& 6.577 &\multirow{2}{*}{4.93(0.96)}&0.76(15)\\
&&&& & 2.844&5.663&&0.89(17)\\[1mm]

\multirow{2}{*}{$^{23}$O$(p,pn)$}&\multirow{2}{*}{--}&\multirow{2}{*}{1.13} &\multirow{2}{*}{5.89} & \multirow{2}{*}{1.00} & 8.765 & 70.474 &\multirow{2}{*}{54.0(10.8)}&0.77(15)\\
&&&& & 8.536&68.636&&0.79(16)\\[1mm]

\multirow{2}{*}{$^{12}$C$(p,2p)$}&\multirow{2}{*}{3.65}&\multirow{2}{*}{0.63} &\multirow{2}{*}{--} & \multirow{2}{*}{--} & 6.143& 26.298 &\multirow{2}{*}{19.2(1.8)[1.2]}&0.73(8)\\
&&&& & 6.498&27.816&&0.69(8)\\

\end{tabular}

\caption{\label{tab:rs3}%
Experimental \cite{Dia18,Ata18,Pan16} and calculated cross sections. The second to fifth columns correspond to the sum of the spectroscopic factors from the prediction of shell-model calculations using WBT interaction for the indicated waves, restricted to bound states of the residual \textit{core}. The next column indicates the single particle cross section $\sigma_\mathrm{sp}$, computed using Dirac (upper value) and PH (lower value) potentials. Next the theoretical cross section $\sigma_\mathrm{th}=\sum C^2S \sigma_\mathrm{sp}$, and the experimental cross sections $\sigma_\mathrm{exp}$ are presented. Finally the quenching factor $R_s=\sigma_\mathrm{exp}/\sigma_\mathrm{th}$ is shown. }
\end{table*}

\medskip
%--------------------------------------------------------------
\section*{Discussion}
%---------------------------------------------------------------
Despite a sizable dispersion of the $R_s$ values shown in Fig.~\ref{fig:Rs}, one observes an overall quenching factor of about 0.7-0.8 with a small, if any, dependence on the asymmetry $\Delta S$. This behaviour is found for the two sets of nucleon--nucleus optical potentials. In order to evaluate the asymmetry dependence, both sets of quenching factors have been fitted with a linear function, which is presented in Fig.~\ref{fig:Rs} along with the results. We obtain from this fit a dependence such as $R_s=0.694(17)-2.5(12)\cdot 10^{-3}\Delta S$ for the calculations employing Dirac potentials and $R_s=0.766(18)+0.4(13)\cdot 10^{-3}\Delta S$ for the calculations with Paris-Hamburg potentials, considering spectroscopic factors from the calculation using WBT interaction. The corresponding reduced $\chi^2$ are, 1.15 and 0.76, for Dirac and Paris-Hamburg calculations, respectively.
%indicate that this linear fits should only be taken as  mere guidance of the overall trend of the calculated $R_s$.  Despite this caveat, 
 The very small value of the slope for both sets of calculations serves as a clear indication of the small dependence of the reduction factors on $\Delta S$. 
 This reduced asymmetry is in qualitative agreement
 %with the analysis performed in \citep{Ata18}, employing the eikonal DWIA theory, and 
 with recent results found for low-energy transfer reactions \citep{Fla13}, but is in contrast with the steep asymmetry found in the systematic study of knockout reactions with heavy targets\citep{Gad08}, represented by the brown band in  Fig.~\ref{fig:Rs}.

Besides the dependence on the optical potentials, the conclusions extracted from our analysis might depend on the  choice of the binding potential and the shell-model interaction. To test the dependence on the former, we have performed calculations with radii obtained with another HF potential, namely, the Skm* interaction.  For weakly-bound nucleons, this leads to an increase in the cross section of about 8$\%$ while for deeply-bound ones the increase is of 5.5$\%$. This result is consistent to that presented in \citep{Gad08}, though it shows a somewhat smaller sensitivity. As in \citep{Gad08}, we find this sensitivity to be small enough so as not to alter the conclusions of this work.

As for the dependence on the shell-model interaction, we have repeated our calculations using Cohen-Kurath and PSDMK interactions as indicated above. The new $R_s$ values are plotted in Fig.~\ref{fig:Rs}(b).
%In Fig.~\ref{fig:Rs}(b) results are also presented using spectroscopic factors from shell-model calculations  
As can be seen, the change in the SF leads to a very small change in the trend, which in this case is of $R_s=0.693(17)-3.1(12)\cdot 10^{-3}\Delta S$ for the calculations employing Dirac potentials and $R_s=0.767(18)-0.3(14)\cdot 10^{-3}\Delta S$ for the calculations with Paris-Hamburg potentials, with reduced $\chi^2$ of 1.29 and 0.86, respectively.

\begin{figure}[tb]
\begin{center}
% {\def\svgwidth{0.9\columnwidth}{\input{zrcoor.pdf_tex}} \par}
 {\centering \resizebox*{\columnwidth}{!}{\includegraphics{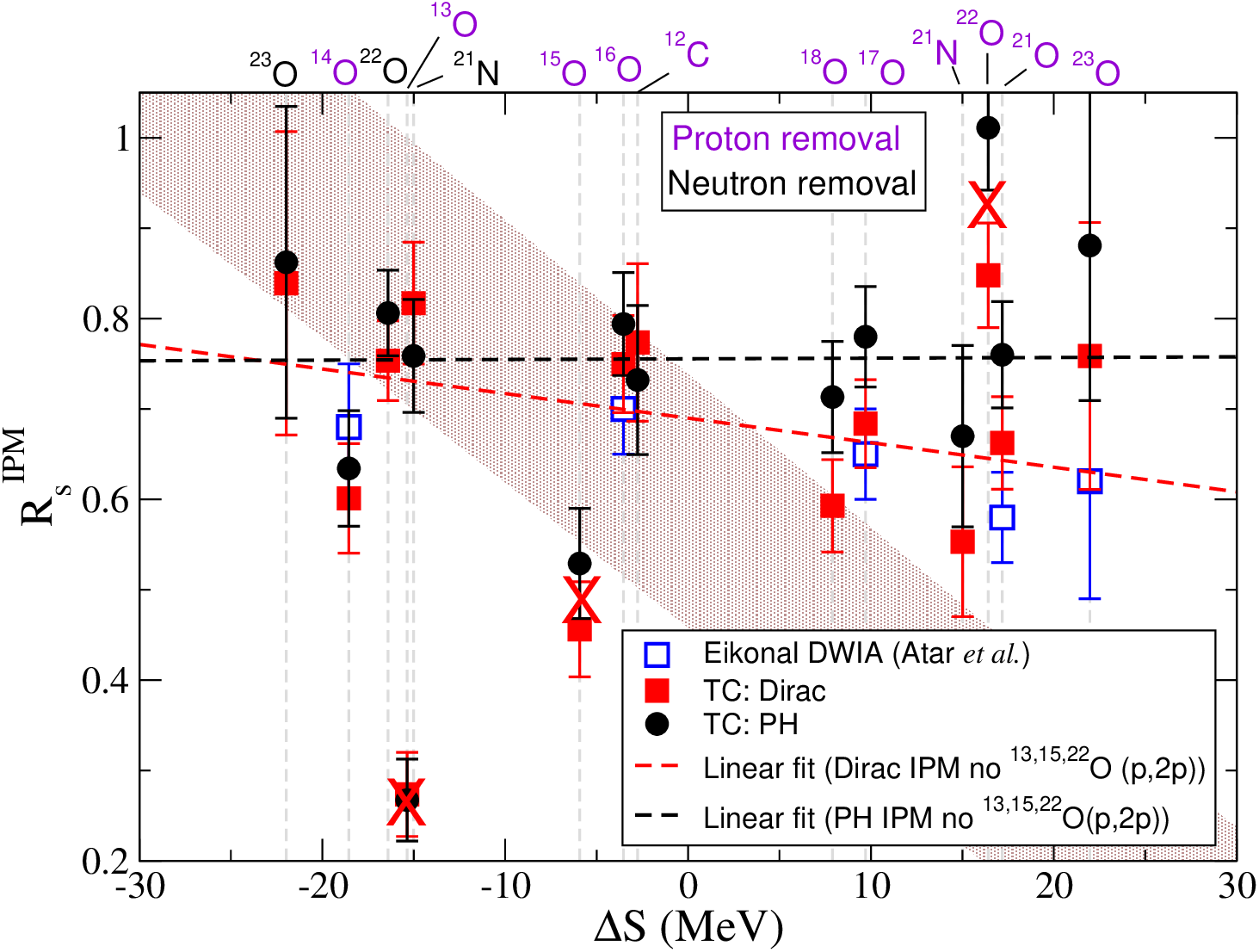}} \par}
\caption{\label{fig:IPM} As Fig.~\ref{fig:Rs}, but presenting calculations where the spectroscopic factors strengths are taken from the IPM. Experimental data from \citep{Dia18} are not shown. Points with red crosses have not been used for the computation of the linear fit.}
\end{center}
\end{figure}

%The extracted $R_s$ values are somewhat larger than those reported in \citep{Ata18}, which oscillate around a mean value of 0.66. 
%Regarding the dependence on $\Delta S$, the present results confirm the weak asymmetry found in the analysis of \cite{Ata18}, based on a selected set of isotopes, and extend these conclusions to the full set of data from  \cite{Ata18}. 
The small dependence on $\Delta$S found in \citep{Ata18} for five selected nuclei is extended by the present results to the analysis of all the isotopes measured in \citep{Ata18}, even though the overall quenching factor found here is somewhat larger than the value of 0.66 from \citep{Ata18}.
%When comparing the reduction factors obtained in this work with those from \citep{Ata18} we find in general an overall smaller quenching (larger $R_s$), centered around a value 0.7-0.8 while those in \citep{Ata18}  oscillate around a mean value of 0.66.
For the cases analyzed in \citep{Ata18}  we find a relatively good agreement for the $R_s$ of $^{14}$O, $^{16}$O and $^{17}$O while there are larger discrepancies for $^{21}$O and $^{23}$O.
We must remark here that in \citep{Ata18} the analysis was performed assuming the Independent Particle Model (IPM), thereby avoiding the introduction of shell-model spectroscopic factors. %As such, comparisons between the calculations in this work and those in \citep{Ata18} are only meaningful when the spectroscopic factors obtained are close to the values obtained in the IPM (6 for $^{16}$O and 2 for the other cases). The spectroscopic factors obtained using WBT interaction (shown in Table \ref{tab:rs3}) are remarkably close to the IPM for $^{14}$O (1.97), $^{16}$O (6.09), $^{17}$O (2.07) and $^{23}$O (1.99), but not for $^{21}$O (1.88). For the latter, if we assume the IPM value for the spectroscopic factor, we get quenching factors of $0.67(5)$ and $0.76(6)$, which are closer (but still larger) than the value of $0.58(4)$ from \citep{Ata18}. 
To have more comparable results we present in Fig.~\ref{fig:IPM} calculations in which the shell-model SF are replaced by the IPM values. We can see in it that the overall results, including the agreement with the values of \citep{Ata18}, tend to follow the same trends as in Fig.~\ref{fig:Rs}, with the notable exceptions of $^{13,15,22}$O$(p,2p)$. Consulting Table \ref{tab:rs3}, we note that for these nuclei the sum of SF lies far from the IPM values (2 or 6), denoting a large fragmentation which sends a large part of the single-particle strength to the continuum of the residual core. As such, we have excluded them from the computation of the overall trend, which yields a result of $R_s=0.692(19)-2.7(14)\cdot 10^{-3}\Delta S$ for the calculations employing Dirac potentials and $R_s=0.755(20)+0.07(15)\cdot 10^{-3}\Delta S$ for the calculations with Paris-Hamburg potentials, with reduced $\chi^2$ of 1.51 and 0.84, respectively. We note that these tendencies are fully compatible to those using SF from shell-model calculations, which we favour since they can consider nuclei with large single-particle-strength fragmentation, as opposed to the IPM, even though the use of shell-model SF may introduce some uncertainties in the calculation.

Our results are in larger disagreement with those presented in \citep{Dia18} (open triangles in Fig.~\ref{fig:Rs}).  For the negative $\Delta$S, the difference is of about 25\% while, for the positive $\Delta$S, our $R_s$ are about twice larger than those from \citep{Dia18}.
%The $R_s$ values reported in that work are significantly smaller and present a larger $\Delta S$ dependence. 
We point out two main reasons  to explain this discrepancy. First, relativistic effects, which are not included in \cite{Dia18}, produce a significant increase of the cross  sections \cite{Yos18}. Second, the different choice for the binding and optical potentials. In particular, the optical potentials adopted here are more absorptive than those used in  \citep{Dia18}, leading to smaller theoretical cross sections and hence larger $R_s$ values. 
It could be argued that the different formalisms used, Faddeev in \citep{Dia18} and Transfer to the Continuum here, may lead also to differences in the results. However, we note that, in a recent benchmark calculation \cite{Yos18}, it was found that Faddeev and Transfer to the Continuum calculations lead to similar $(p,pn)$ cross sections, for the same input ingredients  and using  non-relativistic kinematics.

In Figs.~\ref{fig:Rs} and~\ref{fig:IPM} it can be seen that, overall, the agreement between  TC calculations performed with the two sets of potentials is best for the reactions corresponding to smaller separation energies of the removed nucleon and tends to deteriorate with increasing separation energy. The same can be said for the agreement between the present calculations and those from eikonal DWIA. This can be understood due to the reactions with higher binding energies exploring deeper in the nuclear interior, where distorting potentials are stronger, and thus their effects are more marked in the reaction observables. This implies a greater uncertainty in the interpretation of results for larger binding energies, dependent on the choice of the distorting potentials, which will be less constrained by experimental data for the more exotic species. Even if the particular quenching factors depend on the prescription followed to generate the interaction potentials, their reduced dependence on proton-neutron asymmetry is obtained for both Dirac and Paris-Hamburg calculations, despite their very different origin, and can be established as a solid conclusion from this analysis.

To conclude this section, we note that the weak proton-neutron asymmetry dependence found in this work is also consistent with the conclusions of the recent exclusive $(p,2p)$ measurements of Ref.~\cite{Kaw18} as well as with the state-of-the-art {\it ab-initio} calculations reported in \cite{Ata18} for the proton-hole strength based on the self-consistent Green's function (SCGF) theory \cite{Cip15}. 

\section*{Summary and conclusions}
In summary, we have performed a consistent analysis of all published data from the R$^3$B collaboration to date on total cross sections for $(p,pn)$ and $(p,2p)$ reactions, using the Transfer to the Continuum reaction formalism, focusing on the dependence of the quenching factors on the proton-neutron asymmetry of the studied nuclei. Our analysis yields reduction factors of about 70-80\% with respect to the adopted shell-model spectroscopic factors and with a very small $\Delta S$ dependence.  We have investigated the robustness of these results by performing two analyses using different potential sets, as well as different shell-model interactions. 
Although the calculated $R_s$ values show some dependence on the underlying optical potentials, particularly for the larger binding energies,  the weak $\Delta S$ dependence is consistently observed in both analyses.

The present results agree with those reported in \citep{Ata18}, 
with the results for exclusive $(p,2p)$ measurements reported in \citep{Kaw18} and with those obtained in low-energy transfer reactions \citep{Fla13,Fla18}, although they disagree with the steeper asymmetry found in intermediate-energy nucleon knockout reactions \cite{Gad08}. This discrepancy calls for a revision of the analysis of heavy-ion induced knockout reactions, to clarify how much of the quenching observed and its binding-energy dependence is a sign of strong correlations between nucleons in the nucleus, as has been assumed so far, and how much results from the reaction mechanism for these reactions.

\bigskip 

\section*{Acknowledgements}
We are grateful to Paloma D\'iaz Fern\'andez and Leyla Atar for many useful discussions regarding the experimental data.   
This work has been partially supported by the Spanish
 Ministerio de  Econom\'ia y Competitividad and FEDER funds under project 
 FIS2014-53448-C2-1-P  and by the European Union's Horizon 2020 research and innovation program under grant agreement No.\ 654002. M.~G.-R. acknowledges a research grant from the Spanish Ministerio de Educaci\'on, Cultura y Deporte, Ref: FPU13/04109.

\section*{References}
\bibliography{qfs}

\end{document}